      [1999/12/01 v1.4c Il Nuovo Cimento]
\documentclass{cimento}

\usepackage{graphicx}
\usepackage{graphicx}
\DeclareGraphicsExtensions{.eps,.png}
\usepackage{amsmath}
\usepackage{epsfig}
\usepackage{version}
 \usepackage{mathptmx}  
\usepackage{comment}

\title{Studying the neutron orbital structure by coherent hard exclusive 
processes off
$^3$He}
\author{M.~Rinaldi\from{ins:x}\ETC,
S.~Scopetta\from{ins:y},
}
\instlist{\inst{ins:x} Dipartimento di Fisica, Universit\`{a} degli studi
 di Perugia and
INFN sezione di Perugia, Via Pascoli 06100, Perugia Italy \\
matteo.rinaldi@pg.infn.it
 \inst{ins:y}Dipartimento di Fisica, Universit\`{a} degli studi di Perugia
 and
INFN sezione di Perugia, Via Pascoli 06100, Perugia Italy \\
sergio.scopetta@pg.infn.it}
\PACSes{\PACSit{21.45.}{v}
\PACSit{14.20.}{Dh}
\PACSit{13.60.}{r}}

\begin{document}

\maketitle

\begin{abstract}
Hard exclusive processes, such as Deeply Virtual Compton Scattering (DVCS),
allow to access  generalized parton distributions (GPDs).
By means of an
Impulse Approximation (IA) calculation, it is shown here how, 
in the low momentum transfer 
region, the sum of the GPDs $H$ and $E$, is dominated by the neutron 
contribution. Thanks to this property, $^3$He  could open a new way to 
access the neutron structure information. In this work, a simple and 
efficient extraction procedure of the neutron GPDs, 
able to take into account the nuclear 
effects included in IA analysis, is proposed. 
\end{abstract}

In the last 20 years, many studies have been performed to measure the helicty
quark contributions to the nucleon spin, obtained in DIS or SiDIS experiments,
and the orbital angular momentum 
 (OAM) of the 
partons, crucial steps to solve the so called ``Spin Crisis''. 
Generalized parton distributions \cite{uno} (GPDs), 
which parametrize the non-perturbative 
hadron structure in hard exclusive processes, allow to access important 
information, such as the OAM of the partons inside the nucleon \cite{rassegne}.
The cleanest process to access GPDs is Deeply Virtual Compton Scattering 
(DVCS), \textit{i.e}.
$eH \longmapsto e'H' \gamma$ \ when \ $Q^2\gg M^2$, where $Q^2=-q \cdot q$\ 
is the momentum transfer
between the leptons $e$ and 
$e'$, $\Delta^2$ the one between hadrons $H$ and $H'$ with
momenta $P$ and $P'$, and
$M$ is the nucleon mass.
Another relevant kinematical variable is the
skewedness, $\xi = - \Delta^+/(P^+ + P^{'+})$ 
\footnote{In this paper, $a^{\pm}=(a^0 \pm a^3)/\sqrt{2}$}.
The DVCS cross-section dependence on the GPDs is quite complicated; despite 
of this fact,
experimental data have been obtained and
analyzed from proton and nuclear targets, see refs. \cite{data1, data2}. 
\newpage
\begin{figure}[t]
\vspace{3cm}
\flushleft
\includegraphics{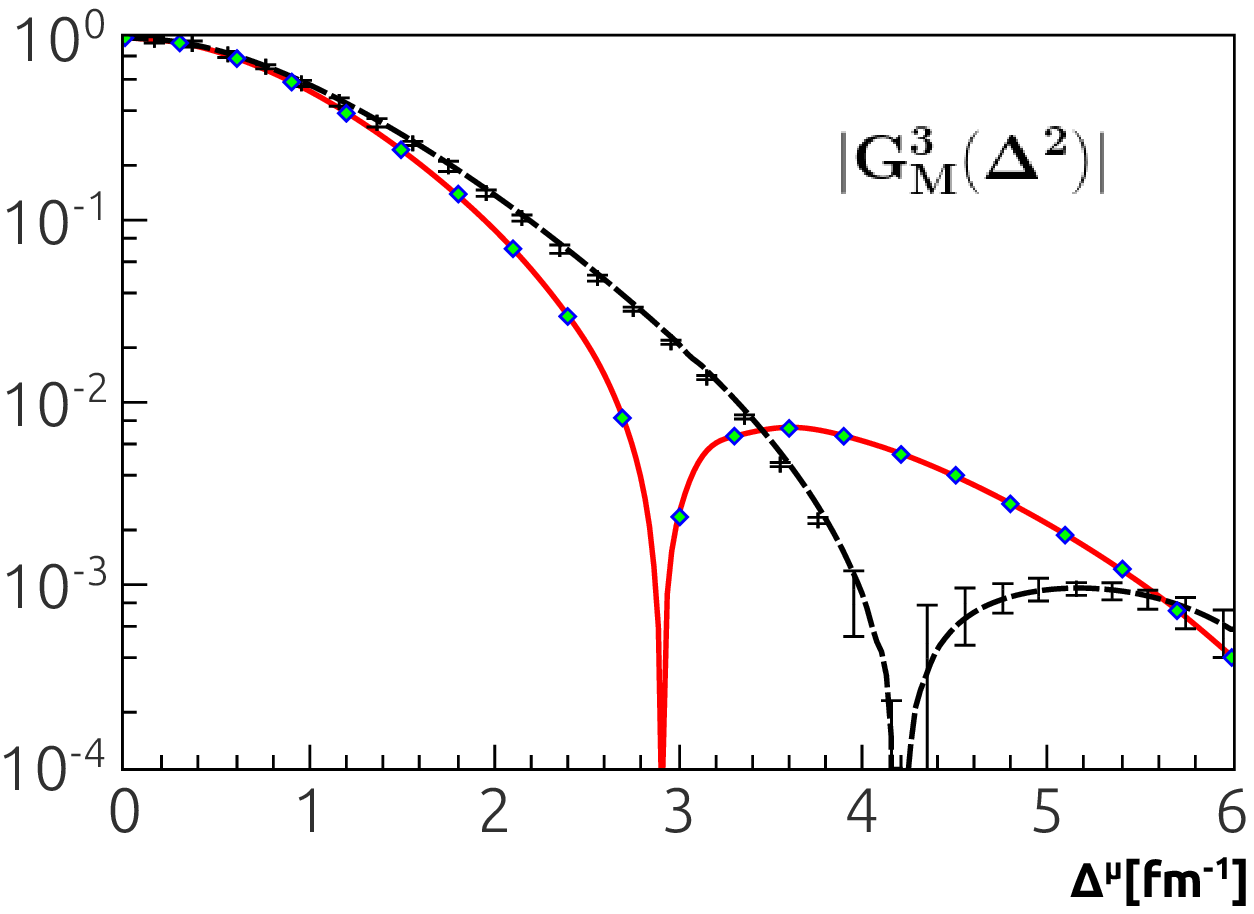}
~~~~~~~~~~~~~~~~~~~~~~~~~~~~~~~~~~~~~~~~~~~~~~~~~~~~~~~~~~~~~~~~~~~~~~
\vskip-3.2cm
\hskip5.9cm
\begin{minipage}{66mm}
{\footnotesize Fig.1: The magnetic ff of $^3$He, $G_M^3(\Delta^2)$, with $\Delta^{\mu} =
\sqrt{-\Delta^2}$. 
Full line: 
the present IA calculation, obtained as the x-integral of $\sum_q \tilde{G}^{3,q}_M$
 (see text). Dashed line: experimental data \cite{dataff}; square points: one-body 
direct calculation, using the Av18 wave function only.
}
\end{minipage}
\end{figure}
\ \ \ \ \  \ \ \ \ \ \ \  \ \ \ \ \ \  \ \ \ \ \ \ 
\vskip5mm
For nuclear targets, the measurement of GPDs is relevant to unveil
medium modifications of bound nucleons and to distinguish between different models,
possibilities excluded {\it e.g.} in DIS experiments.
Nuclear targets are also
necessary to access the neutron information, crucial to obtain,
together
with the proton data, a flavor decomposition of GPDs. 
At this purpose, $^3$He is very promising thanks to its spin structure
(see, \textit{e.g.} ref. \cite{3He}). In fact,
 among the light nuclei, $^3$He is the only one for which the combination
$\tilde{G}_M^{3,q}(x,\Delta^2,\xi) = H^{3}_q(x,\Delta^2,\xi)+
E^{3}_q(x,\Delta^2,\xi)$, whose second moment at $\Delta^2 = 0$ 
gives 
the total angular momentum of the parton of $q$ flavor, 
could be dominated by the neutron contribution.
To this aim $^2$H and $^4$He are not useful, as discussed in ref. \cite{noiarxive} .
To what extent this fact can be used to extract the neutron information,
is shown in refs. \cite{noiarxive,noiold}, and summarized here.
The formalism used in ref. \cite{scopetta}, where a convolution formula for 
GPD $H^3_q$ of
$^3$He was found in IA, has been extended to obtain 
$\tilde{G}_M^{3,q}$:

\begin{eqnarray}
 \tilde G_M^{3,q}(x,\Delta^2,\xi)  = 
\sum_N
\int dE 
\int d\vec{p}~
\tilde{P}^3_N(\vec{p}, \vec{p}',E)
{\xi' \over \xi}
\tilde G_M^{N,q}(x',\Delta^2,\xi'),
\end{eqnarray}

\noindent
where $\tilde{P}^3_N(\vec{p}, \vec{p'},E)$
is a proper combination of components of the spin dependent,
one body off diagonal spectral function:

\begin{eqnarray}
 \label{spectral1}
 P^N_{SS',ss'}(\vec{p},\vec{p}\,',E) 
= 
\dfrac{1}{(2 \pi)^6} 
\dfrac{M\sqrt{ME}}{2} 
\int d\Omega _t
\sum_{\substack{s_t}} \langle\vec{P'}S' | 
\vec{p}\,' s',\vec{t}s_t\rangle_N
\langle \vec{p}s,\vec{t}s_t|\vec{P}S\rangle_N~,
\end{eqnarray}
\noindent
where $x'$ and $\xi'$ are the variables for the bound nucleon 
GPDs, $p \, (p'= p + \Delta)$ and $S,S'(s,s')$
are the 4-momentum and spin projections in the initial (final) state, and
$E= E_{min} +E_R^*$, 
with $E^*_R$ is the excitation energy 
of the two-body recoiling system.
The most important quantity appearing in the definition
eq. (\ref{spectral1}) is
the intrinsic overlap integral
\begin{equation}
\langle \vec{p} ~s,\vec{t} ~s_t|\vec{P}S\rangle_N
=
\int d \vec{y} \, e^{i \vec{p} \cdot \vec{y}}
\langle \chi^{s}_N,
\Psi_t^{s_t}(\vec{x}) | \Psi_3^S(\vec{x}, \vec{y})
 \rangle~
\label{trueover}
\end{equation}
between the wave function
of $^3$He,
$\Psi_3^S$,  
with the final state, described by two wave functions: 
{\sl i)}
the
eigenfunction $\Psi_t^{s_t}$, with eigenvalue
$E = E_{min}+E_R^*$, of the state $s_t$ of the intrinsic
Hamiltonian pertaining to the system of two {\sl interacting}
nucleons with relative momentum $\vec{t}$, 
which can be either
a bound 
or a scattering state, and 
{\sl ii)}
the plane wave representing 
the nucleon $N$ in IA. In order to estimate the nucleon contributions
to  eq. (1),
a numerical evaluation  is needed. To this aim, a simple nucleonic
model of GPDs \cite{Rad1}, which fulfills the main properties, together with the 
overlaps, eq. (3), calculated with the wave function \cite{AV18} of Av18 \cite{pot} 
potential and 
corresponding to the analysis of ref. \cite{overlap}, have been used.
In ref. \cite{scopetta} the main properties of the $^3$He GPD $H^3_q$ 
have been verified,
such as the forward limit and the first moment.
The only possible check for $\tilde G_M^{3,q}$ is its integral 
\begin{figure}[t]

 \vskip-0.5cm  \begin{minipage}{57mm}

 \hskip -4mm \includegraphics[width=57mm]{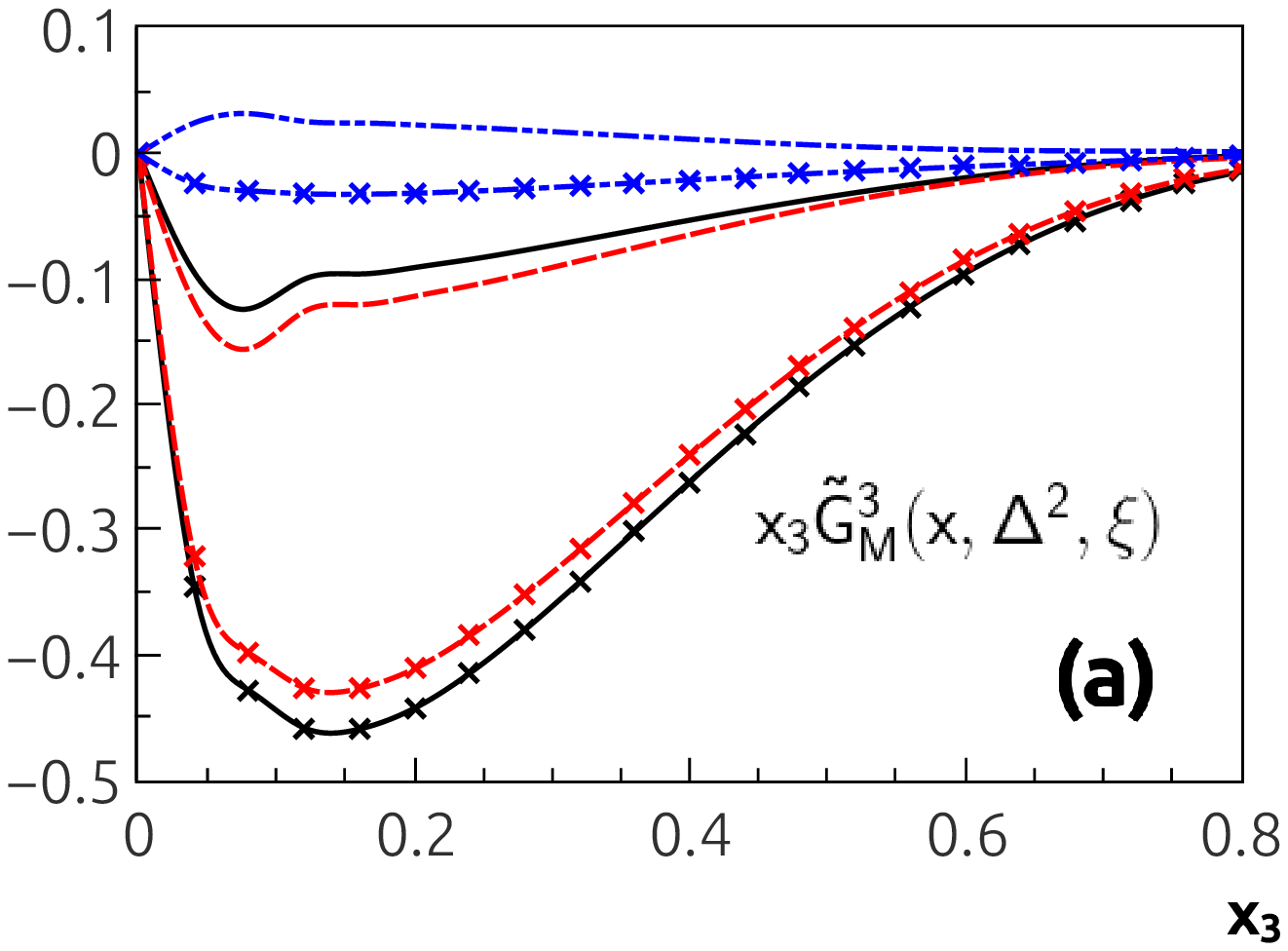}

\end{minipage}

\vskip-5cm   \begin{minipage}{57mm}

 \hskip6.7cm 
\includegraphics[width=57mm]{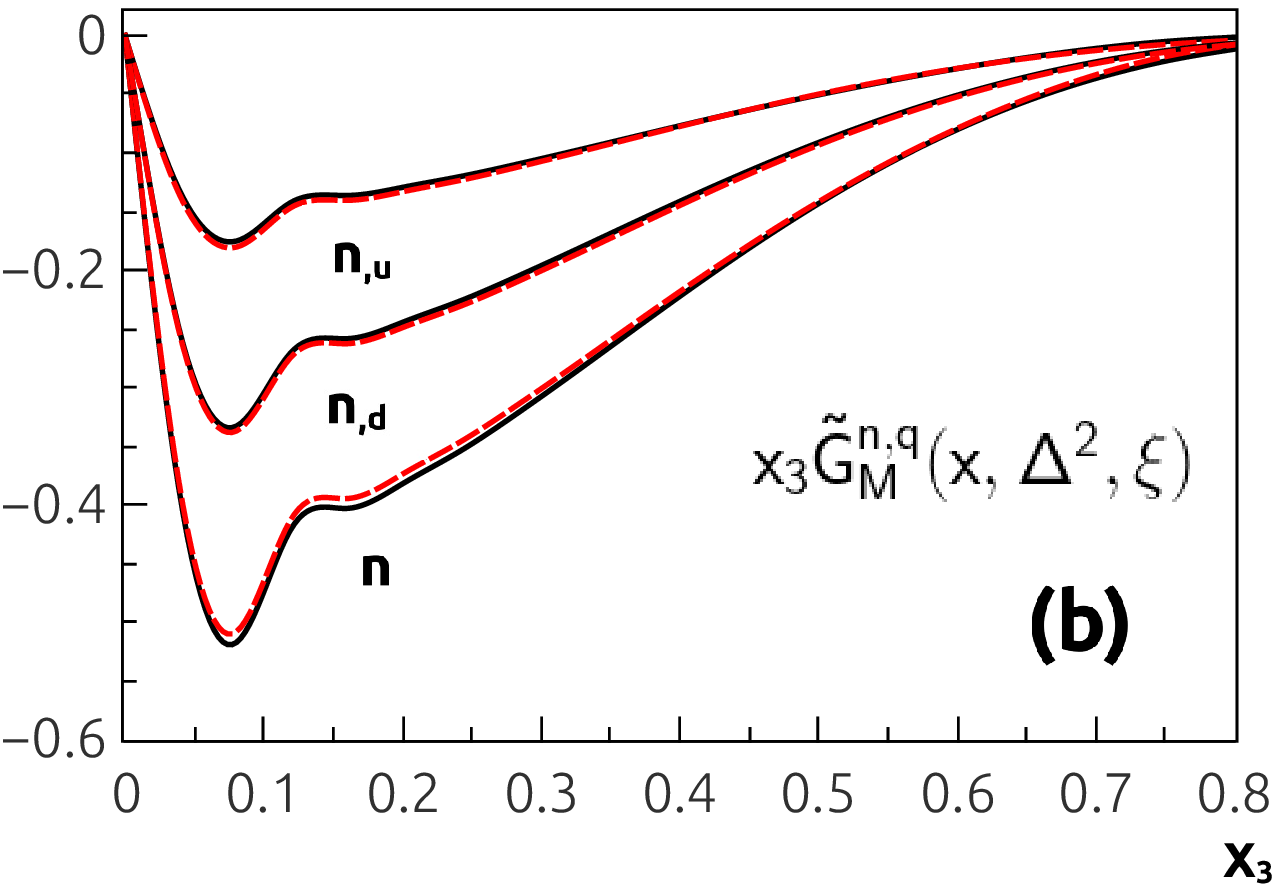}

\end{minipage}

 \vskip 0cm
{\footnotesize Fig.2: (a): The quantity $x_3 \tilde{G}^3_M(x,\Delta^2,\xi)$, where $x_3 =
M_3/M \ x$
and $\xi_3 = M_3/M \ \xi$, shown at $\Delta^2 = 0$ GeV$^2$ and $\xi_3=0$ (stars and lines) and
$\Delta^2 =
-0.1
\ \mbox{GeV}^2$ and $\xi_3=0.1$, \ together with the neutron (dashed) and the proton
(dot-dashed) contribution.
(b): The quantity $x_3 \tilde{G}^{n,q}_M(x,\Delta^2,\xi)$ for the neutron 
at  
$\Delta^2=-0.1 \ \mbox{GeV}^2$ and $\xi_3=0.1$
 with $u$, $d$ and $u+d$ contributions (full lines), compared with the approximation 
 $x_3 \tilde{G}^{n,q,extr}_M(x,\Delta^2,\xi)$, eq. (6), (dashed).}
\end{figure}

\begin{figure}[t]
\vspace{6.5cm}
\includegraphics{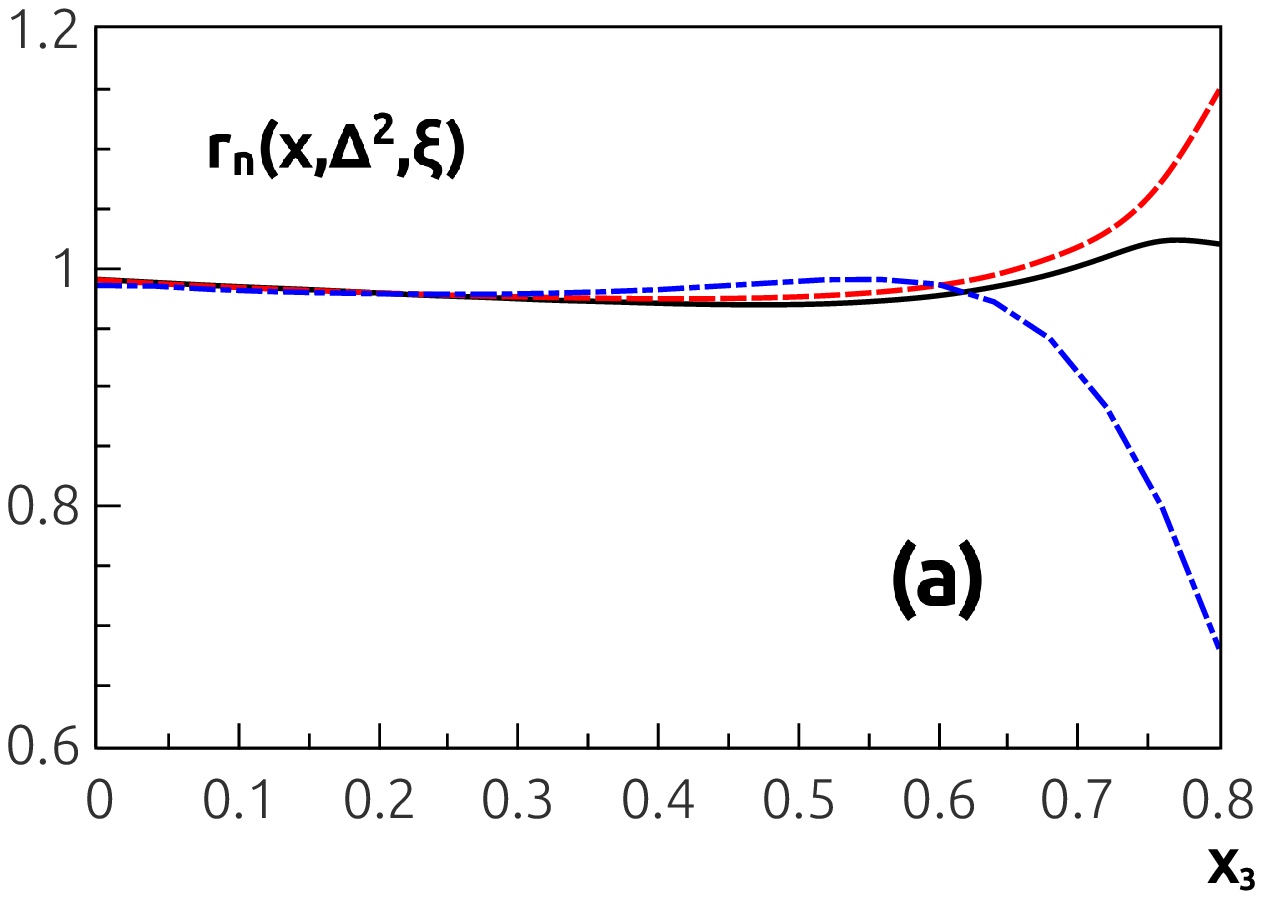}

\includegraphics{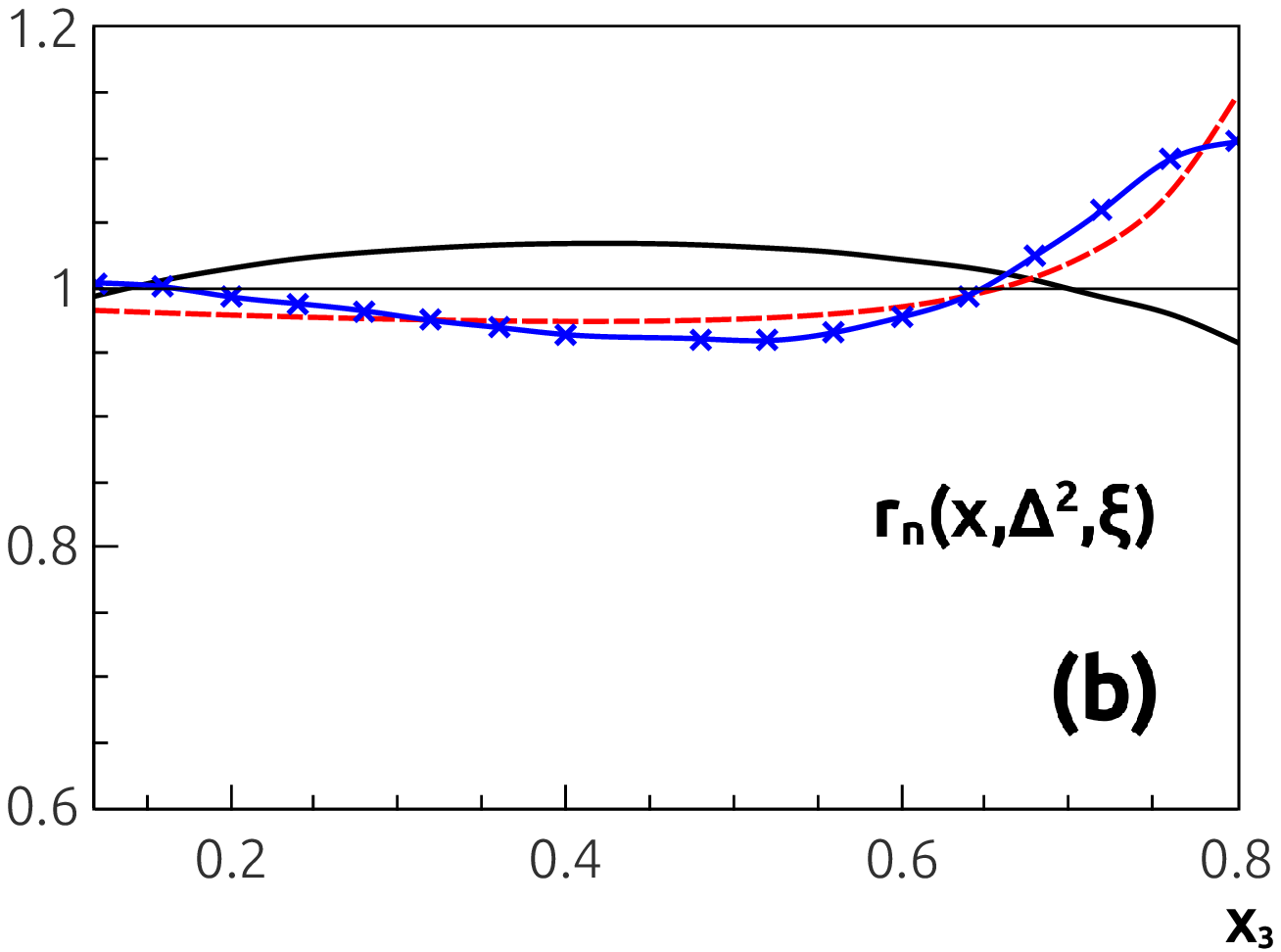}
\vskip-2.2cm
{\footnotesize Fig.3: (a): The ratio $r_n(x, \Delta^2,\xi)= \tilde{G}_M^{n,extr}(x,
\Delta^2,\xi)
/ \tilde{G}_M^{n}(x, \Delta^2,\xi)$, in the forward limit (full), at 
$\Delta^2=-0.1 \ \mbox{GeV}^2$ and $\xi_3=0$ (dashed) and at  
$\Delta^2=-0.1 \ \mbox{GeV}^2$ and $\xi_3=0.1$ (dot-dashed).
(b): $r_n(x, \Delta^2,\xi)= \tilde{G}_M^{n,extr}(x, \Delta^2,\xi)
/ \tilde{G}_M^{n}(x, \Delta^2,\xi)$,
at $\Delta^2$ = 0.1 GeV$^2$ and $\xi_3 = 0$, 
using the model of ref. \cite{Rad1} for the nucleon GPDs (dashed)
, the one of ref. \cite{sv} (full) and the one of ref. \cite{song} (full and stars). }
\end{figure}

 $
 G_M^3(\Delta^2)=\sum_q \int dx \, \tilde G_M^{3,q}(x,\Delta^2,\xi),
$
yielding the magnetic form factor (ff) 
of $^3$He, due the fact that
the forward limit of $ E_M^{3,q}$ is not defined.
Our result is consistent with the Av18
one-body calculation of ref. \cite{schiavilla}.
It is clear
that for $-\Delta^2 \ll 0.15$ GeV$^2$, where DVCS off nuclei can be
performed,
our results compare well also with data \cite{dataff} (see fig.1). 
Now it comes
 the main result of the analysis.
The neutron contribution dominates eq. (1), in particular in the forward
limit, necessary to obtain the OAM; but increasing $-\Delta^2$, the proton
one grows up, see fig.2a, in particular for \textit{u} flavor \cite{noiarxive,noiold}.
Because of
the behavior in $-\Delta^2$  and the complicated convolution formula, an 
extraction procedure of the neutron information is necessary. To this aim, one
can see that
eq. (1) can be written introducing the function 
$g_N^3(z, \Delta^2, \xi )$ 
, the ``light cone spin dependent off-forward momentum distribution'':
\begin{eqnarray}
\tilde G_M^{3,q}(x,\Delta^2,\xi) =   
\sum_N \int_{x_3}^{M_A \over M} { dz \over z}
g_N^3(z, \Delta^2, \xi ) 
\tilde G_M^{N,q} \left( {x \over z},
\Delta^2,
{\xi \over z},
\right)~.
\end{eqnarray}
\noindent
Since $g_N^3(z, \Delta^2, \xi )$ is strongly peaked around $z=1$, with
$x_3 = (M_A/M) x \leq 1$, one has:

\vskip-5.5mm
\begin{eqnarray}
\tilde G_M^{3,q}(x,\Delta^2,\xi) 
& \simeq & {low \,\Delta^2} \simeq   
\sum_N 
\tilde G_M^{N,q} \left( x, \Delta^2, {\xi } \right)
\int_0^{M_A \over M} { dz }
g_N^3(z, \Delta^2, \xi ) 
\nonumber
\\
& = &
G^{3,p,point}_M(\Delta^2) 
\tilde G_M^p(x, \Delta^2,\xi) 
+ 
G^{3,n,point}_M(\Delta^2) 
\tilde G_M^n(x,\Delta^2,\xi)~, 
\end{eqnarray}
\vskip-1mm
\noindent
where the magnetic point like ff has been introduced: 
$G_M^{3,N,point}(\Delta^2)=\int_0^{M_A \over M} dz \, g_N^3(z,\Delta^2,\xi), $ 
which represents the ff of the nucleus if nucleons were point-like particles
with their physical magnetic moments. These quantities are theoretically
well known and their dependence on the nuclear potential is rather weak \cite{noiarxive}.
From eq. (5) the neutron contribution can be extracted:
\begin{eqnarray}
\label{extr}
\tilde G_M^{n,extr}(x, \Delta^2,\xi)  \simeq  
\left\{ \tilde G_M^3(x, \Delta^2,\xi) 
 -  
G^{3,p,point}_{M}(\Delta^2) 
\tilde G_M^p(x, \Delta^2,\xi) \right\}/G^{3,n,point}_M(\Delta^2)~.
\end{eqnarray}
\vskip-3mm

Fig.2b shows our main achievement: the procedure works even beyond the forward
limit, since $G_M^{n,extr}$ 
compares perfectly with  $\tilde G_M^n$, evaluated in the same model
used for the complete calculation of  $\tilde G_M^3$. It is important to 
remark that, for this check, the only theoretically ingredient is the
magnetic point-like ff, which is under control. This crucial result can be analyzed
in details looking at fig.3a, where the ratio
$
r_n (x,\Delta^2,\xi)  = 
{\tilde G_M^{n,extr}(x, \Delta^2,\xi)
/
\tilde G_M^{n}(x, \Delta^2,\xi)}
$
is shown in different kinematical regions and in fig.3b, where $r_n$ is shown using
different nucleonic models in the calculation. 
It is evident that, for $x < 0.7$,
 where data are expected from JLab, the procedure works and  depends weakly
on the nucleonic (see fig.3b and
ref. \cite{noiarxive}) and nuclear models used in the calculation.
In summary, DVCS off $^3$He allows to access 
the neutron information at low $\Delta^2$ and if data were taken at higher 
$\Delta^2$, a relativistic treatment \cite{lussino}
and/or the inclusion of many body currents, beyond the present IA scheme, 
should be implemented.

\end{document}